\RequirePackage{fix-cm}
\RequirePackage{amsmath, bbm}
\documentclass[twocolumn,epjc3]{svjour3}
\pdfoutput=1
\smartqed  % flush right qed marks, e.g. at end of proof
\usepackage{tabularx} % in the preamble
\usepackage{graphicx}
\usepackage{axodraw2}
\usepackage{amscd}
\usepackage{slashed}
\usepackage{multirow}
\usepackage{float}
\usepackage{amsmath}
\usepackage{bm}
\usepackage[caption=false]{subfig}
\journalname{Eur. Phys. J. C}
\begin{document}

\title{$U(1)_{B_3-L_2}$ Explanation of the Neutral Current $B-$Anomalies}


\author{B. C. Allanach\thanksref{addr1}}

\thankstext{e1}{e-mail: B.C.Allanach@damtp.cam.ac.uk}


\institute{DAMTP, University of Cambridge, Wilberforce Road, Cambridge, CB3 0WA, United Kingdom \label{addr1}
}

\date{Received: date / Accepted: date}

\maketitle

\begin{abstract}
We investigate a speculative short-distance force, proposed to explain discrepancies observed   between measurements of certain neutral current decays of $B$ hadrons and their Standard Model predictions. The force derives from a spontaneously broken, gauged $U(1)_{B_3-L_2}$ extension to the Standard Model, where the extra quantum numbers of Standard Model fields are given by third family baryon number minus second family lepton number. The only fields beyond those of the Standard Model are three right-handed neutrinos, a gauge field associated with $U(1)_{B_3-L_2}$ and a Standard Model singlet complex scalar which breaks $U(1)_{B_3-L_2}$, a `flavon'. This simple model, via interactions involving a TeV scale force-carrying $Z^\prime$ vector boson, can successfully explain the neutral current $B-$anomalies whilst accommodating other empirical constraints. In an ansatz for fermion mixing, a combination of up-to-date $B-$anomaly fits, LHC direct $Z^\prime$ search limits and other bounds rule out the domain 0.15 TeV$< M_{Z^\prime} <$ 1.9 TeV at the 95$\%$ confidence level. For more massive $Z^\prime$s, the model possesses a {\em flavonstrahlung}\/ signal, where $pp$ collisions produce a $Z^\prime$ and a flavon, which subsequently decays into two Higgs bosons.  
\end{abstract}

\section{Introduction \label{sec:intro}}
Data from the first decade of running of Large Hadron Collider (LHC)
experiments involving the decays of $B$ hadrons show some 
discrepancies with Standard Model (SM) predictions.
For example, measurements of
the ratio of branching ratios
$R_{K^{(\ast)}}=BR(B \rightarrow K^{(\ast)} \mu^+ \mu^-)/BR(B \rightarrow
K^{(\ast)} e^+ e^-)$~\cite{Aaij:2017vbb,CERN-EP-2019-043}, $BR(B_s \rightarrow \mu^+ \mu^-)$~\cite{Aaboud:2018mst,Chatrchyan:2013bka,CMS:2014xfa,Aaij:2017vad} and some
angular distributions in $K^\ast \mu^+ \mu^-$ decays~\cite{Aaij:2013qta,Aaij:2015oid,Aaboud:2018krd,Sirunyan:2017dhj,Khachatryan:2015isa,Bobeth:2017vxj} all show some
discrepancy (there are others).
Each discrepant observable is
only 1-4$\sigma$ away from SM predictions but collectively,
they point to a roughly similar
conclusion.
Despite a recent flagship LHCb measurement of $R_{K}$
fluctuating somewhat toward its SM prediction (announced at the Moriond
2019 conference), the overall picture remains.
Relative theoretical uncertainties, while taken into
account in the number of sigma, vary from less than 1$\%$ to 20$\%$, 
depending on the particular observable in question.
In summary,
several measurements of $B$ hadron decays are somewhat inconsistent  
with the SM prediction of the 
$(\bar s b)(\bar \mu \mu)$ effective coupling. 
We call these discrepancies the neutral current\footnote{This is to
  distinguish some other discrepancies in $BR(B\rightarrow D^{(\ast)} \tau
  \nu)/BR(B\rightarrow D^{(\ast)} l \nu$~\cite{Zyla:2020zbs},
  which are charged current processes
  and which we do not address.} $B-$anomalies (NCBAs). 

Several different fits to over a hundred $B-$observables~\cite{Alguero:2019ptt,Alok:2019ufo,Ciuchini:2019usw,Aebischer:2019mlg,Datta:2019zca,Kowalska:2019ley,Arbey:2019duh} broadly
agree: they favour a beyond the SM contribution to the weak effective theory operator
\begin{equation}
  {\mathcal L}_{BSM} = -C_9 {\mathcal N}
  (\bar{s} \gamma^\rho P_L b) (\bar \mu \gamma_\rho \mu)+H.c., \label{C9}
\end{equation}
where ${\mathcal N}=1/(36$ TeV$)^2$
(in the present paper, $C_9 \neq 0$ 
 means a contribution {\em beyond}\/ the SM). 
We shall focus on one of the fits for definiteness:
Ref.~\cite{Aebischer:2019mlg}, where the result is that
\begin{equation}C_9  = -0.97  \pm 0.15.\label{expC9}\end{equation}
The coefficient of the operator at the
best-fit point has a pull of 5.9$\sigma$ away from the SM value of 0
(taking the operator
with $P_L$ inserted before the final $\mu$ field in (\ref{C9}) provides an even
better, but comparable, fit, 6.6$\sigma$ away from the SM value). 

One possibility to generate such beyond the SM contributions is from the
interactions of a new electrically neutral, massive, force carrying particle,
dubbed a $Z^\prime$, which has family dependent interactions. In particular,
in order to explain the $B-$anomalies,
the
Lagrangian density should include (with the possible inclusion or exclusion of the second
term) the interaction terms  
\begin{eqnarray}
{\mathcal L}_{int}&=&-  g_{\mu_L}
      \overline{\mu_L} \slashed{Z}^\prime \mu_L - g_{\mu_R} \overline{\mu_R}
      \slashed{Z}^\prime \mu_R \nonumber \\ && - g_{sb} \left(
      \overline{s_L}\slashed{Z}^\prime b_L +H.c.\right), \label{coups}
\end{eqnarray}
where $g_{sb}$, $g_{\mu_L}$ and $g_{\mu_R}$ are all
dimensionless coupling constants. Once the $Z^\prime$ is integrated out, 
in the weak effective field theory, one obtains the Lagrangian density terms
\begin{eqnarray}
{\mathcal L}_{WET}&=&
  -\frac{g_{sb} g_{\mu_L}}{M_{Z^\prime}^2}
  (\overline{s_L} \gamma^\rho b_L) (\overline{\mu_L} \gamma_\rho \mu_L)
  \nonumber \\ &&-
    \frac{g_{sb} g_{\mu_R}}{M_{Z^\prime}^2}
    (\overline{s_L} \gamma^\rho b_L) (\overline{\mu_R} \gamma_\rho \mu_R) +
    H.c. \label{WET}
\end{eqnarray}
These are precisely of the type that can explain the NCBAs: identifying
(\ref{C9}) and (\ref{WET}) we see that
\begin{equation}
C_9=
g_{sb}(g_{\mu_L}+g_{\mu_R}) (36\text{~TeV}/M_{Z^\prime})^2. \label{ident}
\end{equation}

Many models of flavoured $Z^\prime$ vector bosons have been invented based on
spontaneously broken gauged $U(1)$ flavour
symmetries~\cite{Ellis:2017nrp,Allanach:2018vjg}, for example from
$L_\mu-L_\tau$ and other groups~\cite{Gauld:2013qba,Buras:2013dea,Buras:2013qja,Altmannshofer:2014cfa,Buras:2014yna,Crivellin:2015mga,Crivellin:2015lwa,Sierra:2015fma,Crivellin:2015era,Celis:2015ara,Greljo:2015mma,Altmannshofer:2015mqa,Allanach:2015gkd,Falkowski:2015zwa,Chiang:2016qov,Becirevic:2016zri,Boucenna:2016wpr,Boucenna:2016qad,Ko:2017lzd,Alonso:2017bff,Alonso:2017uky,1674-1137-42-3-033104,Bonilla:2017lsq,Bhatia:2017tgo,Ellis:2017nrp,CHEN2018420,Faisel:2017glo,PhysRevD.97.115003,Bian:2017xzg,PhysRevD.97.075035,King:2018fcg,Duan:2018akc,Allanach:2018lvl,Allanach:2018odd,Kang:2019vng,Calibbi:2019lvs,Capdevila:2020rrl,Davighi:2020qqa}.
Some models have several abelian groups in the
extension~\cite{Crivellin:2016ejn}, whilst
some others~\cite{Kamenik:2017tnu,Camargo-Molina:2018cwu,Borah:2020swo}
generate beyond the SM
contributions with loop-level penguin diagrams. 
Some of the models are more ambitious than others, providing more or less
detail toward ultra-violet completion.

In Refs.~\cite{Alonso:2017uky,Bonilla:2017lsq} a gauged $U(1)_{B_3-L_2}$ symmetry was
proposed to 
explain the 
neutral current $B-$anomalies. Both papers are quite detailed in their 
exposition, providing information about fermion mass model
building through additional vector-like representations of the gauge group.

%% More precise model details here
In ref.~\cite{Alonso:2017uky}, Alonso {\em et al}\/ introduce three additional
SM-singlet scalar fields charged under $U(1)_{B_3-L_3}$
and\footnote{$U(1)_{B_3-L_3}$ in the model 
  is equivalent to $U(1)_{B_3-L_2}$ after a change of basis of leptonic
  fields.}
a vector-like fermion for
each Weyl fermion of the SM (plus three right-handed neutrinos). It was shown
how all SM fermion masses and mixings can originate via the Froggatt-Nielsen
mechanism~\cite{Froggatt:1978nt}. The Froggatt-Nielsen mechanism derives effective SM Yukawa
couplings 
from renormalisable interactions which appear to be non-renormalisable from
the low energy effective field theory point of view, once some heavy particles
(the additional heavy vector-like fermions) are integrated out of the
theory. It was shown by Alonso {\em et al}\/ how assumptions about hierarchies
in their masses and further assumptions regarding the renormalisable couplings
of the model translate into empirically feasible values of fermion masses and
fermion mixing.

In Ref.~\cite{Bonilla:2017lsq}, Bonilla {\em et al}\/ introduce two SM-singlet
scalars that are charged under $U(1)_{B_3-L_2}$ along with a second Higgs
doublet, also charged under $U(1)_{B_3-L_2}$. Two vector-like quark doublet
representations are included. Once these scalar fields acquire vacuum
expectation values (with various hierarchies between them assumed),
a realistic pattern of neutrino masses and mixing can be
achieved. In this model, Yukawa couplings leading to quark mixing are obtained
already at the renormalisable level, without invoking the Froggatt-Nielsen mechanism.

The most robustly testable part of the phenomenology of the
models of Alonso {\em et 
  al}\/ and Bonilla {\em et al}\/ is that of the $Z^\prime$, since it is the
interactions of the $Z^\prime$ that explain the NCBAs.
Since the $Z^\prime$ contributes to processes other than those included in the
NCBAs, compatibility with measurements of these other processes then provides constraints upon
each model.  
In both analyses, contributions from the additional
non-$Z^\prime$ states to the NCBAs and to other constraints were neglected by choosing parameters
such that the additional states decouple. 

It is our purpose here to examine the up-to-date collider phenomenology of
similar models 
without worrying about the details of the physics that fixes the fermion mass
data. To this end, we
provide a simplified 
broad-brush formulation of a low energy effective field theory of a gauged,
spontaneously broken
$U(1)_{B_3-L_2}$ model and 
apply the latest bounds and fits, which have changed 
since the original analyses of Refs.~\cite{Alonso:2017uky,Bonilla:2017lsq} due to
a significantly increased integrated luminosity at the LHC\@. 
Since the original analyses, $Z^\prime$ constraints from direct searches have
had an increase 
from 36 fb$^{-1}$ to 139 fb$^{-1}$
and the NCBA fits have changed due to the inclusion of further $B$
meson decay data in some of the observables (the LHCb data set roughly
doubling in size). 
In the effective field
theory, we shall include the $Z^\prime$ boson as well as the flavon,
a complex scalar field whose vacuum expectation value breaks $U(1)_{B_3-L_2}$.
We shall neglect to specify other fields of the model, arguing that
(along similar lines to Alonso {\em et al}\/ and Bonilla {\em et al}\/)  our
analysis
should capture the most pertinent features of the most currently relevant
phenomenology, provided other states are sequestered from it either by
sufficiently weak couplings or heavy masses. Thus, 
compared to Refs.~\cite{Alonso:2017uky,Bonilla:2017lsq}, we do not have as
detailed a model of fermion masses 
but instead we have, in the back of our minds (such as through the
Froggatt-Nielsen mechanism or through mixing with heavy vector-like fermion
representations) an idea of how some small perturbations may be generated
to correct the textures of Yukawa matrices that are predicted at the level of the
renormalisable, unbroken $U(1)_{B_3-L_2}$ theory.
However, 
the latest bounds and fits presented here
will be applicable to all models where
assumptions about the flavour mixing of the
$Z^\prime$ couplings match ours and where other states are sufficiently
decoupled. 

Our paper proceeds as follows: in \S\ref{sec:model}, we
define the effective $U(1)_{B_3-L_2}$ model,
examining the 
$Z^\prime$ couplings to fermions, which are of paramount importance for 
phenomenology.
To specify a model for phenomenological study, it is necessary
to make further assumptions about fermion mixing; these are made in
\S\ref{sec:eg}. Then, in \S\ref{sec:phen}, we examine the current consistency of NCBA
fits with the other experimental constraints.
A novel signal process, {\em
  flavonstrahlung}, is identified. In \S\ref{sec:disc}, we provide a summary and
discussion. Technical definitions of mixing matrices and fields are made
available in~\ref{sec:weakToMass}. 

\section{$B_3-L_2$ Model \label{sec:model}}
The gauge group of the model is
$SU(3) \times SU(2)_L \times U(1)_Y \times
U(1)_{B_3-L_2}$. 
\begin{table}
  \begin{center}
  \begin{tabular}{|cccccc|} \hline
    ${Q_L'}_i$ & ${u_R'}_i$ & ${d_R'}_i$ & $L_1'$ & $L_2' $ & $L_3'$ \\
        0    &     0     &     0     &   0   &  -3    &   0   \\ \hline
    ${e_R'}_1$ & ${e_R'}_2$ & ${e_R'}_3$ & ${\nu_R'}_1$ & ${\nu_R'}_2 $ & ${\nu_R'}_3$ \\
        0     &    -3     &     0     &   0   &  -3    &  0   \\ \hline
    ${Q_L'}_3$ & ${u_R'}_3$ & ${d_R'}_3$  & $H$   & $\theta$ &  \\
        1     &    1      &     1     &  0    &   $q_\theta$    &     \\ \hline          \end{tabular}
  \end{center}
  \caption{\label{tab:assignments} $B_3-L_2$ charge assignments of  fields. A
    prime 
    denotes a weak eigenstate Weyl fermion. 
Under $SU(3) \times SU(2)_L \times U(1)_Y$, the fields have representation
${ Q_L'}_j\sim (3, 2, 1/6)$,
${ L_L'}_j\sim (1, 2, -1/2)$,
${ e_R'}_j\sim (1, 1, -1)$,
${ d_R'}_j\sim (3, 1, -1/3)$,
${ u_R'}_j\sim (3, 1, 2/3)$,
${ \nu_R'}_j\sim (1, 1, 0)$,
$H=(H^+,\ H^0)^T \sim (1,2,1/2) $, respectively.
    $i \in \{
1,2\},\ j \in \{1,2,3\}$ are family indices. The
flavon, $\theta$, is a SM-singlet complex scalar field and $q_\theta$ is a
non-zero rational real number.} 
  \end{table}
We display the charge assignments of the fields in the model under
$U(1)_{B_3-L_2}$ in
Table~\ref{tab:assignments}.
The chiral fermions are all in {\em vector-like}\/ representations with
respect to $U(1)_{B_3-L_2}$ and so standard arguments imply that the symmetry
is free from local perturbative anomalies, given that the SM plus three
right-handed neutrinos is already free of local
gauge anomalies under $SU(3) \times SU(2)_L \times U(1)_Y$. 

At the renormalisable unbroken level, $U(1)_{B_3-L_2}$ predicts that the
Yukawa matrices of SM fermions
(see~\ref{sec:weakToMass} for definitions and conventions)
have the texture
\begin{equation}
  Y_u \sim \left( \begin{array}{ccc}
    \times  & \times &  0 \\
    \times  & \times &  0 \\
    0       & 0      &  \times \\
\end{array}  \right), 
  Y_d \sim \left( \begin{array}{ccc}
    \times  & \times &  0 \\
    \times  & \times &  0 \\
    0       & 0      &  \times \\
  \end{array}  \right),
  Y_e \sim \left( \begin{array}{ccc}
    \times  & 0      &  \times \\
    0       & \times &  0 \\
    \times  & 0      &  \times \\
\end{array}  \right), \label{texture}
\end{equation}
where $\times$ denotes an arbitrary dimensionless entry, which may be non-zero.
From this prediction, we deduce that the Cabbibo-Kobayashi-Maskawa (CKM)
matrix has zero entries for 
$V_{ub}$, 
$V_{cb}$, $V_{ts}$ and $V_{td}$. However, the $U(1)_{B_3-L_2}$ symmetry is
spontaneously broken by the vacuum expectation value
$\langle \theta \rangle$ of a flavon:
a SM singlet scalar $\theta$ with non-zero $B_3-L_2$ charge $q_\theta$. This breaking
will replace the zero entries in (\ref{texture}) by small corrections
generated by non-renormalisable operators.
The model then predicts that the magnitudes of the CKM matrix entries $V_{ub}$,
$V_{cb}$, $V_{ts}$ and $V_{td}$ are suppressed from unity by some small
factor. This 
qualitative expectation~\cite{Alonso:2017uky,Bonilla:2017lsq} agrees with current experimental estimates:
$|V_{cb}|=(41.0\pm1.4) \times 10^{-3}$, $|V_{ub}|=(3.82\pm 0.24) \times 10^{-3}$, 
$|V_{td}|=(8.0\pm 0.3) \times 10^{-3}$, $|V_{ts}|=(38.8\pm 1.1)  \times
10^{-3}$~\cite{Zyla:2020zbs}.
We note that fermion mass data dictate that there should be
hierarchies within the
$\times$ symbols of each matrix in (\ref{texture}).
A more complete
ultra-violet theory could explain such hierarchies.
The (33) entry of each matrix should {\em not}\/ be suppressed, 
in order to explain the hierarchically large masses of
third family fermions as compared to the other two families.
Smaller corrections to the zeroes  will
then indeed predict small entries for the magnitudes of $V_{ub}$,
$V_{cb}$, $V_{ts}$ and $V_{td}$. 

Neutrinos acquire mass through the see-saw mechanism with an initial symmetric
mass matrix (whose basis is defined in (\ref{diracMass})):
\begin{equation}
M_{\nu}= \left(\begin{array}{cccccc}
    0      &    0    &    0    & \dagger &    0    & \dagger \\
    0      &    0    &    0    &    0    & \dagger &    0   \\
    0      &    0    &    0    & \dagger &    0    & \dagger \\
    \dagger &    0   & \dagger &  \ast   &    0    &  \ast  \\
    0      & \dagger &    0    &    0    &    0    &    0   \\
    \dagger &    0   & \dagger &  \ast   &    0    &  \ast  \\
  \end{array}\right), \label{mneut}
  \end{equation}
where the entries marked $\dagger$ are of order the electroweak scale
multiplied by the neutrino Yukawa couplings $Y_\nu$ and we expect the entries
marked $\ast$ to be much greater than  
$\dagger$, since the mass scale $\ast$ is not fixed to the electroweak scale
by any symmetry. As it stands, (\ref{mneut}) has two eigenvalues of order
$\ast$, two of order $\dagger$ and two of order $\dagger^2/\ast$. However, we
expect some of the zeroes in (\ref{mneut}) to be corrected by `small'
non-renormalisable
corrections from the spontaneous breaking of $U(1)_{B_3-L_2}$: in particular,
the bottom right-hand 3 by 3 sub-matrix will be corrected by terms of order
$\ast$ times a small number.
It is expected that such corrections will still
be many orders of magnitude above $\dagger$.
Depending on the value of $q_\theta$, some of the other entries
may be corrected by terms of order $\langle \theta \rangle$.
However, it is not our intention here to
go into the minut\ae\ of fermion mass model building for the model; instead we
shall be content with
the `broad-brush' sketch expected of three very light neutrinos and three
very heavy ones resulting from the expected small corrections and the see-saw mechanism.

We begin with the couplings of the $U(1)_{B_3-L_2}$ gauge boson $Z^\prime_\mu$
to fermions in the Lagrangian in the weak (primed) eigenbasis
\begin{eqnarray}
{\mathcal L}_{Z^\prime \psi} &=& 
-g_{Z^\prime} \left( 
{\overline{{Q_3'}_L}} \slashed{Z}^\prime {Q_3'}_L +
{\overline{{u_3'}_R}} \slashed{Z}^\prime {u_3'}_R +
{\overline{{d_3'}_R}} \slashed{Z}^\prime {d_3'}_R \right. \nonumber \\
&&\left. -3
{\overline{{L_2'}_L}} \slashed{Z}^\prime {L_2'}_L-3
{\overline{{e_2'}_R}} \slashed{Z}^\prime {e_2'}_R-3
{\overline{{\nu_2'}_R}} \slashed{Z}^\prime {\nu_2'}_R
\right), \label{Zpcouplings}
\end{eqnarray}
where
$g_Z^\prime$ is the $U(1)_{B_3-L_2}$ gauge coupling. 
$U(1)_{B_3-L_2}$ is broken by $\langle \theta \rangle \neq 0$ and so the
$Z^\prime$ acquires a mass
\begin{equation}
  M_{Z^\prime} = q_\theta g_F \langle \theta \rangle.
\end{equation}
We shall see below that 
a combination of LHC search bounds and NCBAs
will imply that $M_{Z^\prime}$ is at least
of order the TeV scale.
We assume that the approximately right-handed neutrinos discussed above 
have a much higher mass than 
$M_{Z^\prime}$.
The $Z^\prime$ boson `eats' one real degree of freedom of $\theta$ via the
Brout-Englert-Higgs mechanism~\cite{PhysRevLett.13.321,PhysRevLett.13.508} to form its
longitudinal polarisation mode.
In the spontaneously broken theory, we expand  $\theta = (\langle \theta \rangle + \vartheta)/\sqrt{2}$,
in terms of the one real physical flavon degree 
of freedom, $\vartheta$. Its tree-level mass
$m_{\vartheta}$, 
depends on free parameters 
in the $\theta$ potential, but barring special circumstances
we may expect it to be of order $\langle \theta
\rangle$. 

\begin{figure}
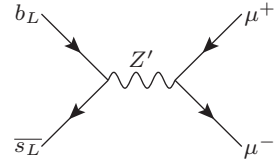

\begin{center}
% Z^\prime diagram
  \begin{axopicture}(80,55)(-5,-5)
    \Line[arrow](0,50)(25,25)
    \Line[arrow](25,25)(0,0)
    \Line[arrow](75,50)(50,25)
    \Line[arrow](50,25)(75,0)
    \Photon(25,25)(50,25){3}{3}
    \Text(37.5,33)[c]{$Z^\prime$}
    \Text(-5,0)[c]{$\overline{s_L}$}
    \Text(-5,50)[c]{$b_L$}
    \Text(82,50)[c]{$\mu^+$}
    \Text(82,0)[c]{$\mu^-$}    
  \end{axopicture}
\caption{\label{fig:NBCAs} Tree-level Feynman diagram of a process
  which contributes to the NCBAs.}
\end{center}
\end{figure}

Writing the weak eigenbasis fermionic fields as 
3-dimensional vectors in family space ${\bf u_R}'$, ${\bf Q_L}'=({\bf
  u_L}',\ {\bf d_L}')$, ${\bf e_R}'$,
${\bf d_R}'$, ${\bf L_L}'=({\bm{\nu}_L}', {\bf e_L}')$, we define the 3 by 3 unitary matrices $V_P$, where
$P \in \{u_R,\ d_L,\ u_L,\ e_R,\ u_R,\ d_R,\ \nu_L,\ e_L \}$.
These transform between the weak
eigenbasis and the mass (unprimed) eigenbasis\footnote{${\bf P}$ and ${\bf P}'$ are column vectors.} as detailed in~\ref{sec:weakToMass}:
\begin{equation}
  {{\bf P}'} = V_P {\bf P}.
  \end{equation}
Re-writing (\ref{Zpcouplings}) in the mass eigenbasis and using the quark and
lepton mixing matrices $V$ and $U$ defined in (\ref{mix})
\begin{eqnarray}
{\mathcal L}_{Z^\prime \psi} &=&-
g_{Z^\prime}  \left(
{\overline{{\bf d_L}}} \Lambda^{(d_L)}_\Xi \slashed{Z}^\prime {\bf d_L} 
+{\overline{{\bf u_L}}} \Lambda^{(u_L)}_\Xi \slashed{Z}^\prime {\bf u_L} 
\right. \nonumber \\  &&+  {\overline{{\bf u_R}}} \Lambda^{(u_R)}_\Xi \slashed{Z}^\prime {\bf u_R} +
{\overline{{\bf d_R}}} \Lambda^{(d_R)}_\Xi \slashed{Z}^\prime {\bf d_R}
 \nonumber \\ &&- 
 3 {\overline{{\bf e_L}}} \Lambda^{(e_L)}_\Omega \slashed{Z}^\prime {\bf e_L} 
-3 {\overline{{\bm{\nu}_L}}} \Lambda^{(\nu_L)}_\Omega \slashed{Z}^\prime
{\bm{\nu}_L} \nonumber \\ &&-
\left. 3 {\overline{{\bf e_R}}} \Lambda^{(e_R)}_\Omega \slashed{Z}^\prime {\bf e_R}
-3 {\overline{{\bm \nu_R}}} \Lambda^{(e_R)}_\Omega \slashed{Z}^\prime {\bm \nu_R}
\right). \label{Zpcoupmass}
\end{eqnarray}
We have 
defined the 3 by 3 dimensionless Hermitian coupling matrices 
\begin{equation}
\Lambda^{(I)}_\alpha := V_{I}^\dagger \alpha V_{I} ,
\label{lambdas}
\end{equation}
where
$I \in \{ u_L, d_L, e_L, \nu_L, u_R, d_R, e_R, \nu_R \}$, $\alpha \in \{ \Xi, \Omega \}$ and
\begin{equation}
\Xi := \left(\begin{array}{ccc}
0 & 0 & 0 \\ 0 & 0 & 0 \\ 0 & 0 & 1 \\
\end{array}\right), \qquad
\Omega := \left(\begin{array}{ccc}
0 & 0 & 0 \\ 0 & 1 & 0 \\ 0 & 0 & 0 \\
\end{array}\right). \qquad
\end{equation}
Provided that $(V_{d_L})_{23} \neq 0$, (\ref{Zpcoupmass})
contains tree-level couplings of the $Z^\prime$ to
$\overline{b_L} s_L$, $\overline{s_L} b_L$ and $\mu^+ \mu^-$.
Thus, it shows promise to explain the NCBAs through processes
such as the one
in Fig.~\ref{fig:NBCAs}.

\section{Example Case \label{sec:eg}}

In order to specify the model further, we should detail the mixing matrices
$V_I$. However, we have not constructed a detailed model for them.
Here, we shall make a simple ansatz for fermion mixing matrices
which is likely to not to be ruled out by
other flavour bounds on flavour changing neutral currents but which
is favourable from the point of view of the NCBAs.
For example, in order to successfully describe the NCBAs, we require
$(V_{d_L})_{23} \neq 0$. 
We shall examine the limit
(which we call `$(B_3-L_2)$eg')
where
\begin{equation}
  V_{d_L} = \left(\begin{array}{ccc}
    1 & 0 & 0 \\
    0 & \cos \theta_{sb} & -\sin \theta_{sb} \\
    0 & \sin \theta_{sb}& \cos \theta_{sb} \\
    \end{array}\right), 
  \end{equation}
$V_{d_R}=1, V_{e_R}=1, V_{e_L}=1$ and $V_{u_R}=1$, meaning that $V_{u_L} = V_{d_L} V^\dag$, $V_{\nu_L}=U^\dag$, where $U$ is the
lepton mixing matrix defined in~\ref{sec:weakToMass}. Thus, the predicted
tree-level flavour changing neutral currents are, aside from the $Z^\prime$ coupling to
$\bar b s$ and $\bar s b$, relegated to the up quarks and neutrinos, where the bounds from experiment are
significantly weaker. 
Our assumptions here are of course strong, but they merely constitute an
example case for phenomenological study in order to assess
viability. 
Extracting the couplings of the $Z^\prime$ relevant for the NCBAs, we have
\begin{equation}
  {\mathcal L} =  -g_{Z^\prime} \left[\left(\frac{1}{2} \sin 2 \theta_{sb} {\bar s}
  \slashed{Z}^\prime P_L b + H.c.\right) - 3{\bar \mu} \slashed{Z}^\prime \mu.
  \right] + \ldots \label{H.c.} \label{eq:op}
  \end{equation}
Thus, by identifying (\ref{eq:op}) with (\ref{ident}), we have
\begin{equation}
  g_{sb}=\frac{g_{Z^\prime}}{2} \sin 2 \theta_{sb}, \qquad
  g_{\mu_L}=g_{\mu_R}=-
\frac{3g_{Z^\prime}}{2}. \label{gs}
\end{equation}
%% Another obvious ansatz to try (the `one-loop' ansatz),
%% would be where we have the $V_I$ as above
%% except for $V_{d_L}=1$. Then, the $(\bar b P_L s)$
%% coupling of the $Z^\prime$ would originate from a one loop diagram involving a $W$
%% boson and a top quark in the loop, \`{a} la~\cite{Camargo-Molina:2018cwu}.
%% In this case, one would have to check whether it is possible to achieve the
%% correct sign of the operator $(\bar b P_L s) (\mu \mu)$ indicated by fits to the
%% NCBAs. 
%% Compared to $(B_3-L_2)$eg, one would have to increase 
%% the $U(1)_{B_3-L_2}$ gauge coupling and/or reduce the $M_{Z^\prime}$ mass 
%% to compensate for the loop factor and explain 
%% the size of effect in the NCBAs. The one-loop ansatz thus more readily falls
%% afoul of current search constraints, 
%% which rule out more strongly coupled and/or lighter,
%% $Z^\prime$ bosons. However,
%% the one-loop ansatz
%% possesses the virtue of being
%% more predictive because it has no free variable $\theta_{sb}$ affecting the
%% phenomenology. We 
%% leave such a study to future work,  adopting the $(B_3-L_2)$eg for now, in
%% order to study its success (or otherwise) in fitting the NCBAs whilst
%% simultaneously satisfying other experimental constraints. 

\section{Phenomenology \label{sec:phen}}
We have now specified the $(B_3-L_2)$eg enough to apply experimental
constraints to it. We first bound its free parameters through the fit to the
NCBAs and then go on to derive other pertinent bounds before considering
predictions.
\subsection{Fit to NCBAs}
\begin{figure}
  \begin{center}
    \unitlength=\columnwidth
    \begin{picture}(1,0.8)(0.2,0)
      \put(0,0){\includegraphics[width=1.4 \columnwidth]{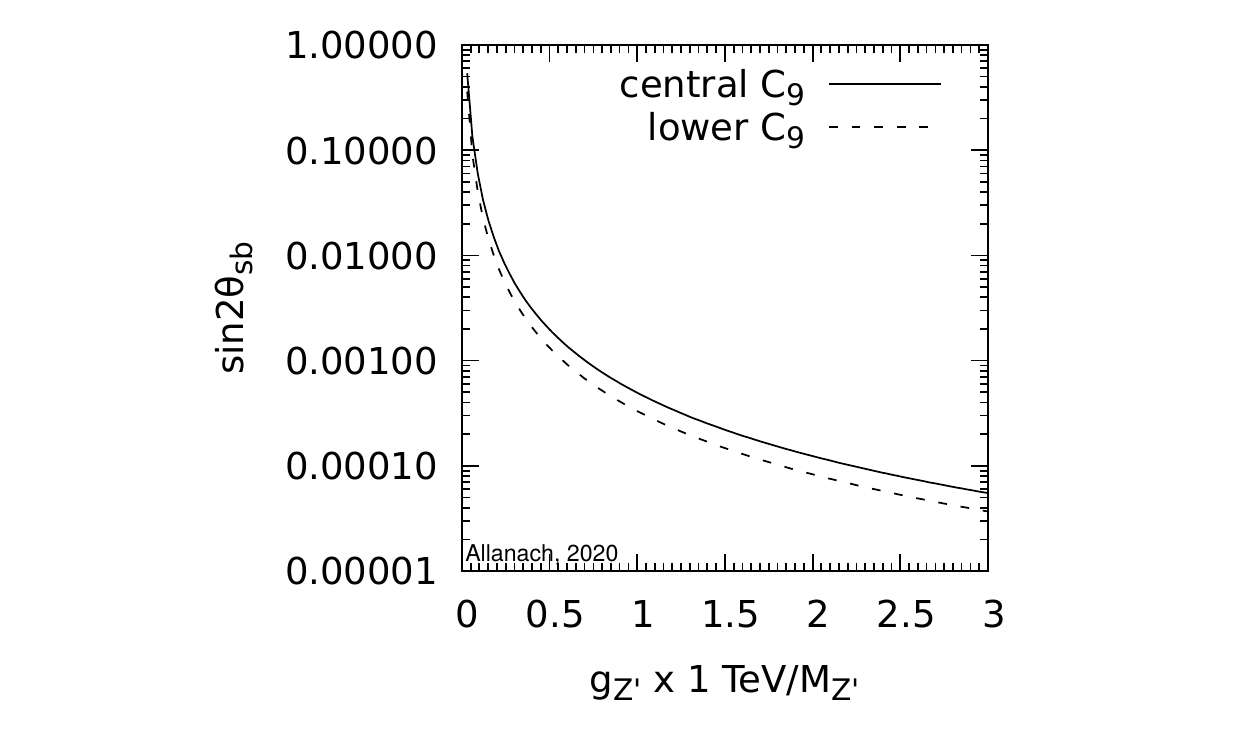}}
      \end{picture}
  \end{center}
  \caption{$\sin 2 \theta_{sb}$ in the $(B_3-L_2)$eg as constrained by a fit~\cite{Aebischer:2019mlg} to
    the NCBAs 
    in (\protect\ref{eq:match}). Central $C_9=-0.97$ and lower $C_9=-0.65$.
    \label{fig:s2b}}
  \end{figure}
At energy scales far below $M_{Z^\prime}$, in the effective theory where the
$Z^\prime$ is integrated out, (\ref{eq:op}) becomes
\begin{equation}
  {\mathcal L} = \frac{3 {g_Z^\prime}^2}{2 M_{Z^\prime}^2} \sin 2 \theta_{sb}
  (\bar{s} \gamma^\rho P_L b) (\bar \mu \gamma_\rho \mu) + H.c., \label{c9op}
\end{equation}
where $\gamma^\rho$ are Dirac matrices, $P_L$ is a left-handed projection matrix and $\rho \in \{0,1,2,3\}$ is a
space-time index.  
The fits prefer no sizeable contributions from the operator 
obtained by switching $P_L \rightarrow P_R$ in
(\ref{c9op})~\cite{Aebischer:2019mlg} and
indeed, since we have assumed $V_{d_R}=1$, we predict none (at tree level).
Substituting $g_{sb}$, $g_{\mu_L}$ and $g_{\mu_R}$ from (\ref{gs}) into
(\ref{ident}), we have
\begin{equation}
\sin 2 \theta_{sb} = -5.1 \times 10^{-4}
\left(\frac{M_{Z^\prime}}{g_{Z^\prime} \text{TeV}}\right)^2 C_9.
  \label{eq:match}
\end{equation}
Requiring that $\sin 2 \theta_{sb} \leq 1$
implies that 
    \begin{equation}
      g_{Z^\prime} \frac{\text{TeV}}{M_{Z^\prime}} \geq 0.023 \sqrt{C_9 / (-0.97)}.
      \label{NCBAbound}
      \end{equation}
    The $(B_3-L_2)$eg has three pertinent free parameters: $M_{Z^\prime}$,
$\theta_{sb}$ and $g_{Z^\prime}$. It will suit us to adopt (\ref{eq:match})
    with the empirically-fitted input for $C_9$
    in order to reduce the number of
free-parameters to two, so that the parameter space of the model can be
captured and plotted in two
dimensions. The central value of $C_9$ as extracted from fits
to the NCBAs shown in (\ref{expC9})
will be the `central $C_9$' value of -0.97, however we will also refer to the
`lower $C_9$' value.
This is the value of $C_9$ which is closest to the SM limit but still fits the relevant
data to within 2$\sigma$ (i.e.\ $C_9=-0.65$~\cite{Aebischer:2019mlg}). 
We display the value of $\sin 2 \theta_{sb}$ for these two cases in
Fig.~\ref{fig:s2b}.

\subsection{$B_s - \overline{B_s}$ mixing}
\begin{figure}
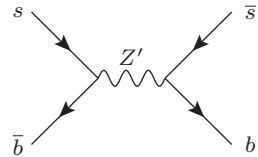

\begin{center}
% Z^\prime diagram
  \begin{axopicture}(80,55)(-5,-5)
    \Line[arrow](0,50)(25,25)
    \Line[arrow](25,25)(0,0)
    \Line[arrow](75,50)(50,25)
    \Line[arrow](50,25)(75,0)
    \Photon(25,25)(50,25){3}{3}
    \Text(37.5,33)[c]{$Z^\prime$}
    \Text(-5,0)[c]{$\overline{b}$}
    \Text(-5,50)[c]{$s$}
    \Text(82,50)[c]{$\overline{s}$}
    \Text(82,0)[c]{$b$}    
  \end{axopicture}
\caption{\label{fig:BsBsbar} Tree-level Feynman diagram of a beyond the SM
  contribution to $B_s - \overline{B_s}$ mixing.}
\end{center}
\end{figure}
Since our $Z^\prime$ couples to bottom and strange (anti-)quarks, it induces a beyond
the SM contribution to $B_s - \overline{B_s}$ mixing via the process in
Fig.~\ref{fig:BsBsbar}.
The value of the bound depends on 
lattice data~\cite{Amhis:2016xyh} which change the SM prediction.
These have
varied significantly over the 
last decade.
We use a recent determination based on lattice data and sum
rules~\cite{King:2019lal} 
which implies that\footnote{In the present paper, we quote all single-sided empirical bounds at the 95$\%$ confidence level.}
$  g_{Z^\prime} \sin 2 \theta_{sb}/2 \leq M_{Z^\prime} / 194 \text{~TeV}$~\cite{Allanach:2019mfl}. Using (\ref{eq:match}), this
implies
the lower bound
\begin{equation}
  g_{Z^\prime} \frac{\text{TeV}}{M_{Z^\prime}} \geq 0.048 \frac{C_9}{-0.97}.
  \label{Bsbound}
  \end{equation}
The fact that this is a {\em lower}\/ bound might at first seem
counter-intuitive, until one realises that, for lower values of $g_{Z^\prime}
\text{TeV}/M_{Z^\prime}$, one can only fit the NCBAs with a larger value of
$\sin 2 \theta_{sb}$, i.e.\ a larger
$Z^\prime$ coupling to bottom and strange (anti-)quarks and therefore a larger
contribution to $B_s-\overline{B_s}$ mixing.

\subsection{Neutrino trident}
\begin{figure}
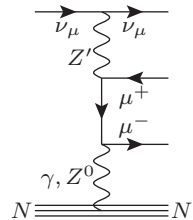

\begin{center}
\begin{axopicture}(55,75)(-5,-2)
  \Line[arrow](0,75)(25,75)
  \Text(12.5,68)[c]{$\nu_\mu$}
  \Line[arrow](25,75)(50,75)
  \Text(37.5,68)[c]{$\nu_\mu$}  
  \Photon(25,75)(25,50){3}{3}
  \Text(17,58)[c]{$Z^\prime$}
  \Line[arrow](50,50)(25,50)
  \Text(37.5,43)[c]{$\mu^+$}
  \Line[arrow](25,50)(25,25)
  \Text(37.5,32)[c]{$\mu^-$}  
  \Line[arrow](25,25)(50,25)
  \Photon(25,25)(25,0){3}{3}
  \Text(12.5,12.5)[c]{$\gamma,Z^0$}
  \Line(0,0)(50,0)
  \Line(0,2)(50,2)
  \Line(0,-2)(50,-2)
  \Text(-5,0)[c]{$N$}
  \Text(55,0)[c]{$N$}
\end{axopicture}
\caption{\label{fig:trident} Tree-level Feynman diagram of a beyond the SM
  contribution to the neutrino trident process.}
\end{center}
\end{figure}
$Z^\prime$s which couple to muon neutrinos contribute to the process $\nu_\mu N \rightarrow
\nu_\mu N \mu^+ \mu^-$, where $N$ is a heavy nucleus, for example by the
process depicted in Fig.~\ref{fig:trident} (there are other diagrams involving
$W$ bosons). In the heavy $Z^\prime$ limit, the
predicted tree-level ratio of
the $(B_3-L_2)$eg model cross-section to the SM one is~\cite{Bonilla:2017lsq} 
\begin{equation}
  \frac{\sigma_{(B_3-L_2)\text{eg}}}{\sigma_{SM}} =
  \frac{1+ (1+4 s_W^2 + 18 v^2 g_{Z^\prime}^2 / M_{Z^\prime}^2)^2}{1 + (1+4
    s_W^2)^2}, \label{trident}
\end{equation}
where $v$ is the SM Higgs vacuum expectation value and $s_W$ is the sine of
the Weinberg angle.
The measurement of the neutrino trident cross section by the CCFR collaboration yields
the constraint
$\sigma_{(B_3-L_2)\text{eg}}/\sigma_{SM}\leq 1.38$~\cite{CCFR}. 
Using the central values $v=246.22$ GeV and
$s_W^2=0.22337$~\cite{Zyla:2020zbs} in (\ref{trident}), this yields
\begin{equation}
  g_{Z^\prime} \frac{\text{TeV}}{M_{Z^\prime}} \leq 0.62. \label{numTrident}
  \end{equation}

\subsection{Anomalous Magnetic Moment of the Muon}
\begin{figure}
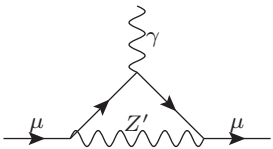

  \begin{center}
    \begin{axopicture}(100,50)
      \Photon(50,50)(50,25){3}{3}
      \Text(56,37.5)[c]{$\gamma$}
      \Line[arrow](75,0)(100,0)
      \Text(87.5,5)[c]{$\mu$}
      \Line[arrow](50,25)(75,0)
      \Line[arrow](25,0)(50,25)
      \Line[arrow](0,0)(25,0)
      \Text(12.5,5)[c]{$\mu$}
      \Photon(25,0)(75,0){3}{6}
      \Text(50,6)[c]{$Z^\prime$}
      \end{axopicture}
    \caption{\label{fig:g-2} Leading new physics contribution to the anomalous
      magnetic moment of the muon.}
  \end{center}
\end{figure}
The $U(1)_{B_3-L_2}$eg has the potential to explain measurements of
the anomalous magnetic moment of the muon $(g-2)_\mu$, which disagrees with SM
predictions. Current estimates for the
discrepancy are~\cite{Zyla:2020zbs}
\begin{equation}
  \Delta a_\mu= \frac{\Delta(g-2)_\mu}{2} =
  (26.1 \pm 6.3 \pm 4.8) \times 10^{-10}, \label{amu}
\end{equation}
where the first uncertainty quoted is experimental and the second theoretical. Adding the
uncertainties in quadrature implies that there is a 3.3$\sigma$ discrepancy
between SM predictions and the experimental measurement. 
A leading contribution to $\Delta a_\mu$ from the $Z^\prime$ is depicted in
Fig.~\ref{fig:g-2}. In total, the $Z^\prime$ corrections yield~\cite{Pospelov:2008zw}
\begin{equation}
  \Delta a_\mu((B_3-L_2)\text{eg}) =
  \frac{3 g_{Z^\prime}^2}{4 \pi^2} \left[
  \frac{m_\mu^2}{M_{Z^\prime}^2} +
  \mathcal{O}\left(\frac{m_\mu^4}{M_{Z^\prime}^4} \right) \right].  \label{gm2pred}
\end{equation}
Equating (\ref{amu}),(\ref{gm2pred}), we find that in order to fit the
anomalous magnetic moment at the 2$\sigma$ level, 
\begin{equation}
1.1  <g_{Z^\prime} \frac{\text{TeV}}{M_{Z^\prime}}< 2.2.
\end{equation}
Comparing with (\ref{numTrident}), we see that the 2$\sigma$ region preferred
by measurements of the anomalous magnetic moment of the muon are in tension
with CCFR measurements of the neutrino trident process.

\subsection{$Z^0 \rightarrow \mu^+ \mu^- Z^\prime$}
\begin{figure}
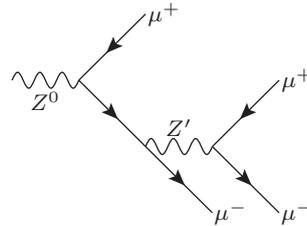

\begin{center}
\begin{axopicture}(110,75)
  \Photon(0,50)(25,50){3}{3}
  \Text(12.5,42)[c]{$Z^0$}  
  \Line[arrow](50,75)(25,50)
  \Text(57,75)[c]{$\mu^+$}  
  \Line[arrow](25,50)(50,25)
  \Line[arrow](50,25)(75,0)
  \Line[arrow](100,50)(75,25)
  \Line[arrow](75,25)(100,0)
  \Photon(50,25)(75,25){3}{3}  
  \Text(62.5,32)[c]{$Z^\prime$}
  \Text(82,0)[c]{$\mu^-$}  
  \Text(107,50)[c]{$\mu^+$}
  \Text(107,0)[c]{$\mu^-$}  
\end{axopicture}
\caption{\label{fig:z4l} Tree-level diagram of a new physics contribution to
  $Z\rightarrow 4 \mu$ decay.}
\end{center}
\end{figure}
For $Z^\prime$ particles whose mass is less than that of the $Z^0$ boson,
i.e.\ $M_{Z^\prime}<M_Z$, a recent CMS search in 77.3 fb$^{-1}$ of 13 TeV $pp$
collisions for $Z^0 \rightarrow \mu^+ \mu^- Z^\prime$ followed by $Z^\prime
\rightarrow \mu^+ \mu^-$ provides constraints~\cite{Sirunyan:2018nnz}.
Stringent 95$\%$ upper bounds upon the
product of branching ratios $$R_{4\mu}=BR(Z^0 \rightarrow \mu^+ \mu^- Z^\prime)\times BR(Z^\prime
\rightarrow \mu^+ \mu^-)$$ at the level of $10^{-7}-10^{-8}$
are presented as a function of $M_{Z^\prime} \in [5
\text{~GeV},\ 70 \text{~GeV}]$. For a given value of $M_{Z^\prime}$ in the
aforementioned range  and a
reference value of $g_{Z^\prime}$, we use
     {\tt MadGraph\_2\_6\_5}~\cite{Alwall:2014hca} to calculate the value of
     $R_{4\mu}$. In the parameter range considered for which this decay
     is relevant, $BR(Z^\prime
     \rightarrow \mu^+ \mu^-)$ is independent of $g_{Z^\prime}$ to a very good
     approximation and so $R_{4\mu}$ is predicted to be
     proportional to $g_{Z^\prime}^2$. We can thus scale $g_{Z^\prime}$ to
     find its value at the
     95$\%$ upper limit. 

\subsection{$Z^\prime$ width and perturbativity}
The partial width of a $Z^\prime$ decaying into a Weyl fermion $f_i$
and Weyl anti-fermion ${\bar f}_j$ is
\begin{equation}
\Gamma_{ij} = \frac{C}{24 \pi} |g_{ij}|^2
M_{Z^\prime}, \label{eq:partial}
\end{equation}
where $g_{ij}$ is the coupling of the $Z^\prime$ boson to $f_i \bar f_j$ and
$C$ is the number of colour degrees of freedom of the fermions 
(here, 3 or 1). In the limit that $m_t / M_{Z^\prime} \rightarrow 0$, we may
approximate all fermions as being massless.
Summing over fermion species (it
is simplest to do this in the weak eigenbasis), 
we obtain a total width $\Gamma$:
\begin{equation}
  \frac{\Gamma}{M_{Z^\prime}} = \frac{13 g_{Z^\prime}^2}{8 \pi}. \label{eq:total}
\end{equation}
To remain in the perturbative r\'e{gime} such that we may trust our perturbative
calculations, we should have
$\Gamma/M_{Z^\prime}<1$, i.e. $g_{Z^\prime} < \sqrt{8
  \pi/13}=1.4$. Substituting this into (\ref{Bsbound}) yields an upper bound
$M_{Z^\prime} \leq 29 (-0.97/C_9)$ TeV
from perturbativity, fits to NCBAs and $B_s-\overline{B_s}$ mixing
measurements.  

\subsection{LHC $Z^\prime$ Searches}
\begin{figure}
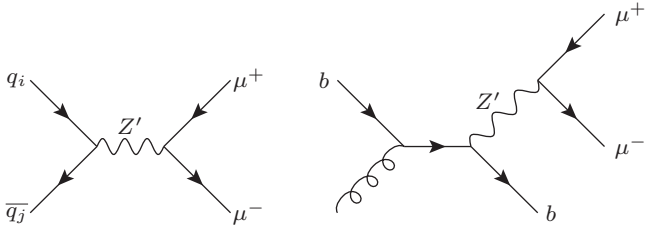

\begin{center}
% Z^\prime diagram
\begin{axopicture}(180,75)(-20,0)
\Line[arrow](-40,50)(-15,25)
\Line[arrow](-15,25)(-40,0)
\Line[arrow](35,50)(10,25)
\Line[arrow](10,25)(35,0)

\Gluon(75,0)(100,25){3}{3}
\Line[arrow](75,50)(100,25)
\Line[arrow](100,25)(125,25)
\Line[arrow](125,25)(150,0)
\Photon(125,25)(150,50){3}{3}
\Line[arrow](175,75)(150,50)
\Line[arrow](150,50)(175,25)
\Text(70,50)[c]{$b$}
\Text(155,0)[c]{$b$}
\Text(185,75)[c]{$\mu^+$}
\Text(185,25)[c]{$\mu^-$}
\Text(131.5,42.5)[c]{$Z^\prime$}

\Photon(-15,25)(10,25){3}{3}
\Text(-2.5,33)[c]{$Z^\prime$}
\Text(-45,50)[c]{${q_i}$}
\Text(-45,0)[c]{$\overline{{q_j}}$}
\Text(42,50)[c]{$\mu^+$}
\Text(42,0)[c]{$\mu^-$}
\end{axopicture}
\end{center}
\caption{Example Feynman diagrams of tree-level $B_3-L_2$ inclusive
  $Z^\prime$ production at the
  LHC followed by decay into muons.
  $q_{i,j} \in \{ u,c,d,s,b \}$ are such that the 
  combination ${q_i} \overline{{q_j}}$
  has zero electric
  charge. \label{fig:prod}}    
\end{figure}
The ATLAS experiment has performed various searches in $pp$ collisions at the LHC for
resonant $Z^\prime$
vector bosons decaying into different final states. None of them have found a
significant signal to date and so lower limits are placed upon the production
cross-sections times branching ratio as a function of the invariant mass of
the final state. For example, a 36.1 fb$^{-1}$ 13 TeV search in $t \bar t$
imposes $\sigma \times BR(Z^\prime \rightarrow t \bar t)<10$
fb for large $M_{Z^\prime}$~\cite{Aaboud:2018mjh,Aaboud:2019roo}. A di-tau
final state search from the 8 TeV run imposes $\sigma  \times BR(Z^\prime \rightarrow \tau^+
\tau^-)<3$ fb for large $M_{Z^\prime}$~\cite{Aad:2015osa}. However, the most
constraining channel to date for the $(B_3-L_2)$eg is from a $Z^\prime \rightarrow
\mu^+\mu^-$ in 139 fb$^{-1}$ of 13 TeV $p p$ collisions~\cite{Aad:2019fac},
where, for $M_{Z^\prime}=6$ TeV, $\sigma \times BR(Z^\prime \rightarrow \mu^+
\mu^-)<0.015$ fb, where $\sigma$ is the fiducial $Z^\prime$ production cross-section.
We shall therefore use this search to constrain the model\footnote{The
  analogous CMS search has yet to be published}. 
Feynman diagrams of example
$Z^\prime$ production
signal processes are shown in Fig.~\ref{fig:prod}. 

In this ATLAS di-muon resonance search, each muon is required to have a transverse momentum $p_T>30$
GeV, pseudo-rapidity magnitude $|\eta|<2.5$ and a di-muon invariant mass
$m_{\mu\mu}>225$ GeV. ATLAS has already taken efficiencies into account in
their published bounds 
so there is no need to simulate the detector. Upper bounds $s(M_{Z^\prime},
z)$ on $\sigma \times BR(Z^\prime \rightarrow \mu^+ \mu^-)$
are published for 
$z:=\Gamma/M_{Z^\prime}$ values from 0 to 0.1~\cite{atlasData}. In
Ref.~\cite{Allanach:2019mfl}, it was shown that 
a function
\begin{equation}
  s(z,M_{Z^\prime}) = s(0,M_{Z^\prime})   \left[    \frac{s(0.1,M_{Z^\prime})}{s(0,M_{Z^\prime})}
    \right]^{\frac{z}{0.1}}
  \label{lim}
  \end{equation}
fits the given published bounds well in the given domain
$z \in [0,0.1]$. We
shall also use (\ref{lim}) to extrapolate slightly outside of this domain, but
will delineate regions of parameter space where the bound is extrapolated
rather than interpolated.

The $(B_3-L_2)$eg model was encoded into {\tt UFO} format via
\texttt{FeynRules}~\cite{Degrande:2011ua,Alloul:2013bka} for inclusion into
  an event generator. We calculate the fiducial cross-section
  $\sigma(pp \rightarrow Z^\prime \rightarrow \mu^+ \mu^-$
  with
  the {\tt MadGraph\_2\_6\_5}~\cite{Alwall:2014hca} event
  generator for a centre of mass
  energy of 13 TeV.
  We have added the possibility of producing an additional
  jet along with the $Z^\prime$ so that the second diagram of
  Fig.~\ref{fig:prod} is included in our estimate of the
  cross-section. 
 We also use five flavour
  parton distribution functions to re-sum initial state $b-$quark
  logarithms~\cite{Lim:2016wjo} and neglect interference with SM
  backgrounds. 
  \begin{table}
    \begin{center}
      \begin{tabular}{|c|c|} \hline
        $M_{Z^\prime}$ & 3 TeV \\
        $g_{Z^\prime}$ & 0.15 \\
        $\sin 2 \theta_{sb}$ & 0.20 \\ \hline
        $\Gamma$ & 35 GeV \\
        $s(M_{Z^\prime}, z)$ & 0.069 fb \\
        $\sigma(p p \rightarrow Z^\prime \rightarrow \mu^+ \mu^-)\ (+j)$ & 0.033 fb
        \\
        $BR(Z^\prime\rightarrow \mu^+ \mu^-)$ & 0.46 \\
        $BR(Z^\prime\rightarrow t \bar t)$ & 0.15 \\
        $BR(Z^\prime\rightarrow b \bar b)$ & 0.15 \\                
        $\sigma(b \bar b \rightarrow Z^\prime \rightarrow \mu^+ \mu^-)$ & 0.026 fb
        \\
        $\sigma(g b \rightarrow Z^\prime b \rightarrow \mu^+ \mu^- b)$ &
        0.007 fb \\
        $\sigma(s \bar b \rightarrow Z^\prime \rightarrow \mu^+ \mu^-)$ &
        $6.1\times 10^{-4}$ fb\\
        \hline        \end{tabular}
    \end{center}
    \caption{\label{tab:point}
      Example point in $(B_3-L_2)$eg parameter space
      that fits the NBCAs (for central $C_9=-0.97$)
      and survives all constraints.
      We show the largest partonic 
      contributions to the cross-section at the bottom of the table.
       For the
      last two rows, the $CP$ conjugated process has been added to the
      cross-section contribution. `$(+j)$' refers to the fact that the
      cross-section includes the addition of another jet in the final
      state.} 
    \end{table}
We display an allowed parameter space point ($M_{Z^\prime}=3$ TeV, $g_{Z^\prime}=0.15$)
in
Table~\ref{tab:point}. From the table, we can see that the dominant process is
$b \bar b \rightarrow Z^\prime\rightarrow \mu^+ \mu^-$, the sub-dominant process
is ($b g \rightarrow Z^\prime b \rightarrow \mu^+ \mu^- b$ plus the 
$CP$ conjugated process).
The other 
tree-level processes  simulated
make a negligible contribution to the
cross-section. 
  
\begin{figure}
  \begin{center}
    \unitlength=\columnwidth
    \begin{picture}(1,0.8)(0.25,0.05)
      \put(0,0){\includegraphics[width=1.4 \columnwidth]{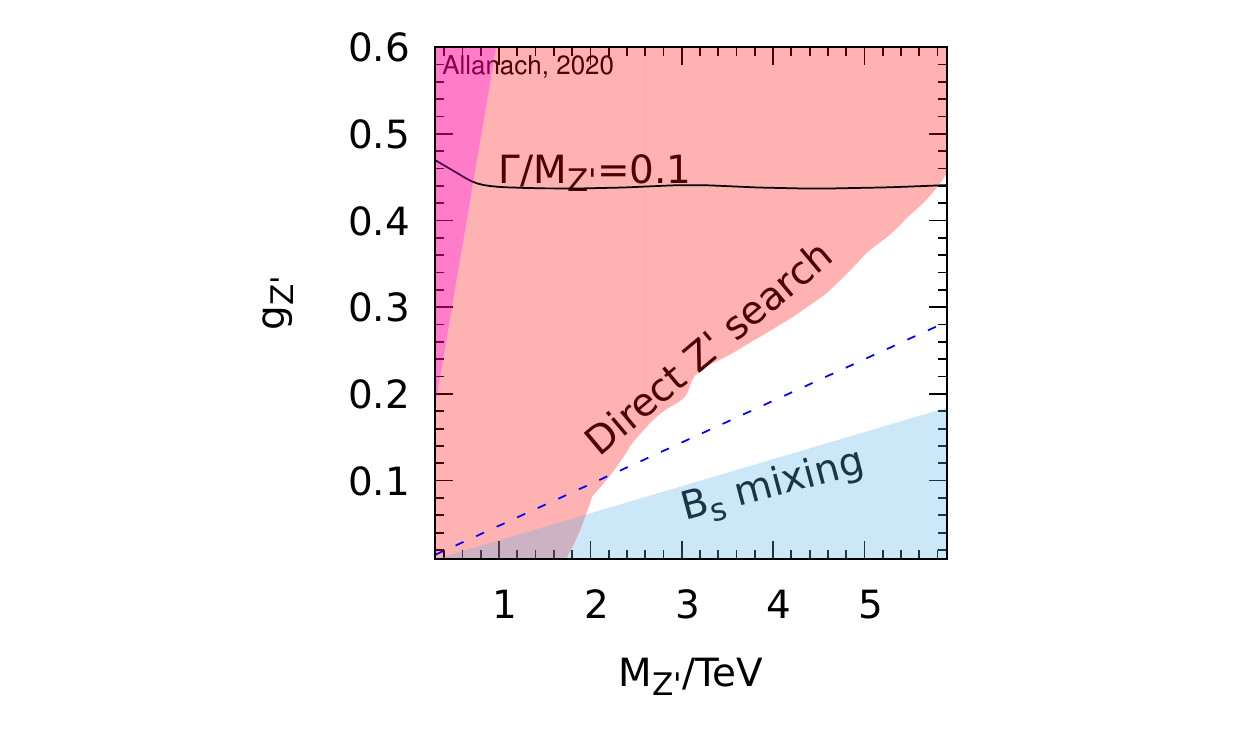}}
      \end{picture}
  \end{center}
  \caption{\label{fig:lower} Constraints upon $(B_3-L_2)$eg for
    $M_{Z^\prime}>300$ GeV.
    $\sin 2\theta_{sb}$ has been set as in (\ref{eq:match}) such
    that every point 
    fits the NCBAs. The white region is currently allowed.
    The red and blue coloured regions show the 95$\%$ excluded regions from
    a 13 TeV 139 fb$^{-1}$ 
    ATLAS $Z^\prime \rightarrow \mu^+ \mu^-$ search~\cite{Aad:2019fac} and from
    $B_s-\overline{B_s}$ mixing as in (\protect\ref{Bsbound}),
    respectively. The latter bound moves from the 
    blue coloured region at lower $C_9=-0.65$ to the region below the dashed
    line for central $C_9=-0.97$. The magenta region in the top left-hand
    corner shows the region ruled
    out by the neutrino trident process.
    The direct search bound is extrapolated above the solid curve and
    interpolated between ATLAS data below it, according to (\protect\ref{lim}). }
  \end{figure}
In Fig.~\ref{fig:lower}, we display constraints upon the $(B_3-L_2)$eg
parameter space for $M_{Z^\prime}>300$ GeV. 
There is only a small region of parameter space where the ATLAS
di-muon resonance search bounds have been extrapolated (slightly): above the
solid curve. 
The white region of the figure is allowed by all constraints. We see that
$M_{Z^\prime}>1.9$ TeV from these. The direct search constraint does not
change by eye from the one shown in the figure when one chooses the central value
of $C_9=-0.97$ from the NCBA fit or the lower value.
We may understand this by the fact that $\sin 2 \theta_{sb}$ is small
throughout the vast majority of the plot, whichever value of $C_9$ is used, 
in accordance with Fig.~\ref{fig:s2b}. The
dominant $Z^\prime$ production amplitude
is proportional to the $Z^\prime \bar b b$
coupling, which is proportional to
$g_{Z^\prime} \cos 2 \theta_{sb}\approx g_{Z^\prime}$ and so loses
the sensitivity\footnote{(\ref{eq:partial}) and (\ref{eq:total}) show that $BR(Z^\prime \rightarrow \mu^+
  \mu^-)$ has no dependence on $C_9$ through $\sin 2\theta_{sb}$ either.}
that $\sin 2 \theta_{sb}$ has on $C_9$ through (\ref{eq:match}). 
The $B_s$ mixing bound {\em is}\/ however sensitive
to a change in $C_9$ (via its effect on $g_{sb}$)
and the bound becomes the dashed
line for central $C_9$. So: for central $C_9$, one concludes that
$M_{Z^\prime}>2.2$ TeV.
For either value of $C_9$ and throughout the allowed parameter space shown,
$BR(Z^\prime \rightarrow \mu^+ \mu^-)$, 
$BR(Z^\prime \rightarrow \bar b b)$ 
and $BR(Z^\prime \rightarrow \bar t t)$ do not change (to the significant
figure quoted) from the values in Table~\ref{tab:point}.

\begin{figure}
  \begin{center}
    \unitlength \columnwidth
    \begin{picture}(1,0.8)(0.25,0.05)
      \put(0,0){\includegraphics[width=1.4 \columnwidth]{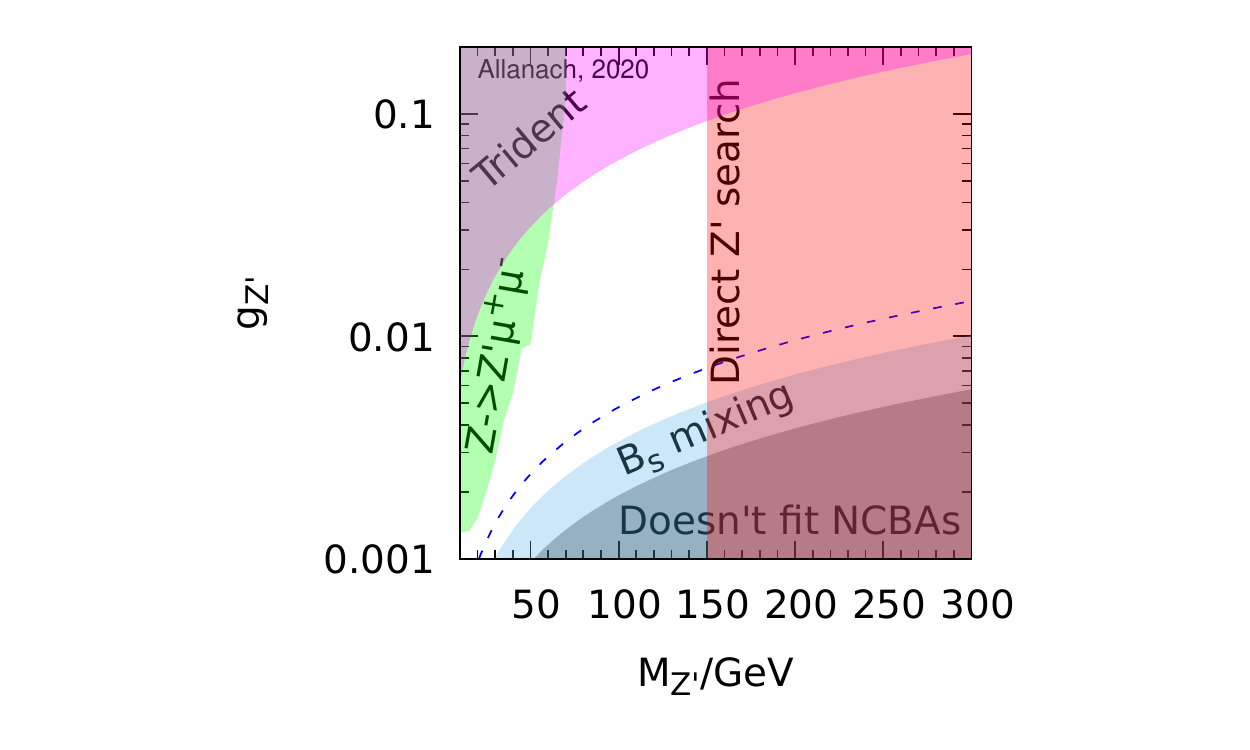}}
      \end{picture}
  \end{center}
  \caption{\label{fig:light} Constraints upon $(B_3-L_2)$eg for
    $M_{Z^\prime}\leq 300$ GeV.
    $\sin 2\theta_{sb}$ has been set as in (\protect\ref{eq:match}) such
    that every point 
    fits the NCBAs. The white region is currently allowed.
    The red and blue coloured regions show the 95$\%$ excluded regions from
    a 13 TeV 36.1 fb$^{-1}$ 
    ATLAS $Z^\prime \rightarrow \mu^+ \mu^-$ search~\cite{Aad:2019fac} and from
    $B_s-\overline{B_s}$ mixing as in (\protect\ref{Bsbound}),
    respectively. The latter bound moves from the 
    blue coloured region at lower $C_9=-0.65$ to the region below the dashed
    line for central $C_9=-0.97$. The magenta region shows the region ruled
    out by the neutrino trident process and the green region by a CMS search in
    77.3 fb$^{-1}$ of 13 TeV $pp$
    collisions for $Z^0 \rightarrow \mu^+ \mu^- Z^\prime\rightarrow 4
    \mu$~\cite{Sirunyan:2018nnz}. In the grey region at the bottom of the
    plot, the effect of the $Z^\prime$ on the NCBAs is too weak to fit them.}
\end{figure}
Bonilla {\em et al}\/ showed that, in the model of Ref.~\cite{Bonilla:2017lsq},
a region of parameter space with 
$M_{Z^\prime}<$300 GeV could pass all constraints. We now re-examine this
lighter mass range;
our analysis closely follows that of
Bonilla {\em et al}, except for the fact that we constrain the parameter space
to always fit the NCBAs and that we have updated the
constraints from $Z \rightarrow \mu^+ \mu^- Z^\prime \rightarrow 4 \mu$ with a
new search from CMS.  

The ATLAS search for $Z^\prime \rightarrow
\mu^+\mu^-$ used above provided no constraints for $M_{Z^\prime} \leq 300$ GeV. Thus, we
have used an earlier ATLAS search\footnote{A similar CMS search exists~\cite{Sirunyan:2018nnz}, but
  doesn't reach quite as low values of $M_{Z^\prime}$ and so we do not use it.}
for $Z^\prime \rightarrow \mu^+ \mu^-$ in 36.1$^{-1}$ fb of 13 TeV $pp$
collisions, which presented exclusions  on
generic\footnote{Since in the relevant part of the parameter space, our
  $Z^\prime$s are predicted to be very narrow, we use the bounds for the
  narrowest $Z^\prime$s given, i.e.\ $\Gamma_{Z^\prime}/M_{Z^\prime}=0.02$.}
$Z^\prime$s
for $M_{Z^\prime} \geq 150$ GeV~\cite{Aaboud:2017buh}. We calculate the
acceptance (for muon transverse momenta
$p_T>30$ GeV and muon pseudorapidities $|\eta|<2.5$) times cross
section times branching ratio for $Z^\prime\rightarrow \mu^+ \mu^- (+ j)$
using {\tt MadGraph\_2\_6\_5}. We find that the
region 150 GeV$< M_{Z^\prime} <$300 GeV is excluded by this search for the
entire domain
$0.001\leq g_{Z^\prime} \leq 0.2$. The available parameter space is shown in
Fig.~\ref{fig:light}. We see that the $(B_3-L_2)$eg may fit the NCBAs for
$M_{Z^\prime}<150$ GeV while still passing other experimental constraints. 

\subsection{Flavonstrahlung \label{sec:flav}}
\begin{figure}
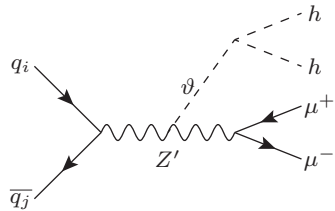

\begin{center}
% Z^\prime diagram
\begin{axopicture}(70,70)(-25,0)
\Line[arrow](-40,50)(-15,25)
\Line[arrow](-15,25)(-40,0)
\Line[dash](10,25)(35,60)
\Line[dash](35,60)(60,70)
\Line[dash](35,60)(60,50)
\Line[arrow](60,35)(35,25)
\Line[arrow](35,25)(60,15)
\Photon(-15,25)(35,25){3}{6}
\Text(10,15)[c]{$Z^\prime$}
\Text(17.5,42.5)[c]{$\vartheta$}
\Text(-45,50)[c]{${q_i}$}
\Text(-45,0)[c]{$\overline{{q_j}}$}
\Text(67,35)[c]{$\mu^+$}
\Text(67,15)[c]{$\mu^-$}
\Text(65,70)[c]{$h$}
\Text(65,50)[c]{$h$}
\end{axopicture}
\end{center}
\caption{Feynman diagram of {\em flavonstrahlung}\/ process at a hadron
  collider. 
  $q_{i,j} \in \{ u,c,d,s,b \}$ are such that the 
  combination ${q_i} \overline{{q_j}}$
  has zero electric
  charge. \label{fig:flavonstrahlung}}    
\end{figure}
In the unbroken $U(1)_{B_3-L_2}$ theory, $\theta$ 
interacts with the Higgs boson via the Lagrangian density term
$-\lambda_{\theta H} \theta \theta^\dag H H^\dag$.
Supposing that the dimensionless
coefficient $\lambda_{\theta H} \neq 0$, the flavon
$\vartheta$ will then decay into
two physical Higgs bosons $h h$ with approximately 100$\%$ branching ratio.
Moreover, the $\theta$ kinetic term leads to the Lagrangian density term
$g_{Z^\prime}^2 q_\theta^2 Z^{\prime}_\mu {Z^{\prime}}^\mu \langle \theta
\rangle \vartheta$ after spontaneous symmetry breaking. 
Thus, if a proton-proton collider has sufficient energy and luminosity, it may
produce $Z^\prime \vartheta$, leading to the spectacular signature of $\mu^+
\mu^- h h$, where $\mu^+ \mu^-$ have a resonance at an invariant mass of
$M_{Z^\prime}$ and
$hh$ have one at the flavon mass $m_{\vartheta}$.
This `flavonstrahlung' process is
depicted in Fig.~\ref{fig:flavonstrahlung}.
Flavonstrahlung would probably not be the first detection of beyond the
SM physics in the model: $Z^\prime$ production followed
by decay into $\mu^+\mu^-$ would most likely be the first, followed perhaps by
$Z^\prime \rightarrow t \bar t$ and $b \bar b$. Flavonstrahlung is
suppressed 
compared to exclusive $Z^\prime$ production because of its larger
final-state phase space and kinematics, and would thus require significantly more
luminosity and partonic energy to detect.

\section{Discussion \label{sec:disc}}
Spontaneously broken $U(1)_{B_3-L_2}$~\cite{Alonso:2017uky,Bonilla:2017lsq}
has parameter space that is consistent 
with contemporary direct search limits whilst fitting neutral current
$B-$anomalies and 
passing other indirect bounds (the most constraining being 
those from
measurements of $B_s-\overline{B_s}$ mixing).

We have provided a simple broad-brush formulation of the $U(1)_{B_3-L_2}$
model, similar  
to the one of the Third Family Hypercharge Model (TFHM)~\cite{Allanach:2018lvl} and
variants~\cite{Allanach:2019iiy}.
We then
presented an example case for phenomenological study, the `$(B_3-L_2)$eg'.
Fig.~\ref{fig:lower} shows that in the $(B_2-L_2)$eg the current empirical constraints imply
that $M_{Z^\prime}>1.9$ TeV or $M_{Z^\prime}<150$ GeV. In the latter lighter
region, one may obtain significant corrections to the anomalous magnetic
moment of the muon (however, neutrino trident constraints are in 2$\sigma$ tension with
the current 2$\sigma$-preferred region of $(g-2)_\mu$).
The fact that $M_{Z^\prime}<150$ GeV is currently allowed motivates further
effort in order to push interpretations of LHC $\mu^+ \mu^-$ resonance 
searches to
lower invariant masses where, admittedly, backgrounds are steeply increasing. 

The constraints in Figs.~\ref{fig:lower},\ref{fig:light} apply to any gauged,
spontaneously broken $U(1)_{B_3-L_2}$ model where our assumptions about the
$Z^\prime$ couplings detailed in \S\ref{sec:eg} approximately hold.
This is
the case for the model of Alonso {\em et al}~\cite{Alonso:2017uky}, which
found the weaker bound of 
$M_{Z^\prime} > 1.0$ TeV from the various predecessor constraint data and NCBA
fits.
The bound has moved to $M_{Z^\prime} > 1.9$ TeV with the latest fits and
data. Alonso {\em et al}\/ did not consider $M_{Z^\prime}<300$ GeV, but our
results in Fig.~\ref{fig:light} show that $M_{Z^\prime}<150$ GeV is currently viable. 
The model of Bonilla {\em et al}~\cite{Bonilla:2017lsq} does not match
the pattern of $(B_3-L_2)$eg $Z^\prime$ couplings to left-handed down quarks
and so our results are not directly applicable to it.

The direct $Z^\prime \rightarrow \mu^+ \mu^-$ search constraints on
the $(B_3-L_2)$eg in Fig.~\ref{fig:lower} are comparable 
to those on similarly constructed TFHMeg models\footnote{In
  Refs.~\cite{Allanach:2018lvl,Allanach:2019iiy}, the associated production
  process $Z^\prime j$ was not included, however.}.
In TFHMs though, 
the Higgs doublet is necessarily charged 
under the additional $U(1)$ in order to allow a renormalisable top Yukawa
coupling (which seems necessary, given that it is of order 1 and so
is inconsistent with a small effective coupling induced by symmetry breaking).
This leads to tree-level $Z-Z^\prime$ mixing, associated strong bounds
from inferences of the $\rho$ parameter~\cite{Davighi:2020nhv}: indeed, these
entirely disallow the $M_{Z^\prime}\leq 300$ GeV region for the TFHMs. 
The $U(1)_{B_3-L_2}$ model is not subject to these 
strong bounds, however, since the SM Higgs doublet is uncharged under
$U(1)_{B_3-L_2}$.

In \S\ref{sec:flav}, we have identified a novel flavonstrahlung signal
process, where $pp$ collisions result in
$Z^\prime$ plus flavon production, followed by $Z^\prime$ decay into $\mu^+ \mu^-$ and flavon decay into $hh$. This process will also be present in
other similar NCBA-explaining $U(1)$ extensions which are broken by a SM
singlet, since the 
flavon field used to break the $U(1)$ extension
will generically have couplings with the SM Higgs doublet. Thus, for example,
TFHMs also predict the possibility of flavonstrahlung.

\section*{Acknowledgements}
This work has been partially supported by STFC Consolidated HEP grants
ST/P000681/1 and ST/T000694/1. We thank other members of the
Cambridge Pheno Working Group and W.\ Murray for discussions.

\appendix
\section{Conventions and fermion mixing\label{sec:weakToMass}}
Here, we detail the rotation of
fermion fields to the mass basis in order to
fix our conventions.
We write 
\begin{eqnarray}
{\bf u_L'}&=&\left( \begin{array}{c} u_L' \\ c_L' \\ t_L' \\ \end{array}
\right), \qquad
{\bf d_L'}=\left( \begin{array}{c} d_L' \\ s_L' \\ b_L' \\ \end{array} \right), \qquad
{\bm \nu_L'}=\left( \begin{array}{c} {\nu_e'}_L \\ {\nu_\mu'}_L
  \\ {\nu_\tau'}_L \\ \end{array} \right),
\nonumber \\
{\bf e_L'}&=&\left( \begin{array}{c} e_L' \\ \mu_L' \\ \tau_L' \\ \end{array}
\right),\qquad
{\bf u_R'}=\left( \begin{array}{c} u_R' \\ c_R' \\ t_R' \\ \end{array}
\right), \qquad
{\bf d_R'}=\left( \begin{array}{c} d_R' \\ s_R' \\ b_R' \\ \end{array}
\right),
\nonumber \\
{\bf e_R'}&=&\left( \begin{array}{c} e_R' \\ \mu_R' \\ \tau_R' \\ \end{array}
\right), \qquad
{\bm \nu_R'}=\left( \begin{array}{c} {\nu_e'}_R \\ {\nu_\mu'}_R
  \\ {\nu_\tau'}_R \\ \end{array} \right),
\end{eqnarray}
along with the SM fermionic electroweak doublets
\begin{equation}
{\bf Q_L'}_i=\left( \begin{array}{c} {\bf u_L'}_i \\ {\bf d_L'}_i \end{array}
\right),\qquad
{\bf L_L'}_i=\left( \begin{array}{c} {\bm \nu_L'}_i \\ {\bf e_L'}_i \end{array}
\right).  
\end{equation}

The fermions acquire their masses through the terms
\begin{eqnarray}
-\mathcal{L}_{Y}&=&\overline{\bf Q'_L}  Y_u \tilde H {\bf u'_R} +
\overline{\bf Q'_L}  Y_d H  {\bf d'_R} +
\overline{\bf L'_L}  Y_e H  {\bf e'_R} + \nonumber \\ &&
\overline{\bf L'_L} Y_\nu \tilde H {\bm \nu'_R} + H.c. +
\frac{1}{2}{\overline{\bm \nu'_R}^c} M
         {\bm \nu_R'}, 
 \label{yuk}
\end{eqnarray}
where $Y_u$, $Y_d$ and $Y_e$ are dimensionless complex coupling constants,
each written as a 3 by 3 matrix in family space.
The matrix $M$ is a 3 by 3 complex symmetric matrix of mass dimension
1, ${}^c$ denotes the
charge conjugate of a field and $\tilde H = ({H^0}^\ast, -H^-)^T$.
After electroweak symmetry breaking and the $W^\pm$ boson eating the
electrically charged
components of the Higgs doublet, we may write
$H=(0,\ (v + h)/\sqrt{2})$,
where 
$h$ is the physical Higgs
boson field
and (\ref{yuk}) includes the fermion mass terms 
\begin{eqnarray}
-\mathcal{L}_{Y}&=&\overline{\bf u'_L} V_{u_L} V_{u_L}^\dagger m_u V_{u_R}
V_{u_R}^\dagger {\bf u'_R} + \nonumber \\ &&
\overline{\bf d'_L} V_{d_L} V_{d_L}^\dagger m_d  V_{d_R} 
V_{d_R}^\dagger {\bf d'_R} + \nonumber \\ &&
\overline{\bf e'_L} V_{e_L} V_{e_L}^\dagger m_e  V_{e_R} 
V_{e_R}^\dagger {\bf e'_R} + \nonumber \\ &&
\frac{1}{2} ( {\overline{\bm \nu_L'}}\ \overline{{\bm \nu_R'}^c}) M_\nu
\left( \begin{array}{c} {\bm {\nu_L'}}^c \\ {\bm \nu_R'} \\
\end{array}
  \right)
  +H.c. \nonumber \\
  && + \ldots, \label{diracMass}
\end{eqnarray}
where
\begin{equation}
M_\nu = \left( \begin{array}{cc} 0 & m_{\nu_D} \\
  m_{\nu_D}^T & M \\ \end{array} \right),
\end{equation}
$V_{I_L}$ and $V_{I_R}$ are 3 by 3 unitary mixing matrices for each
species $I$, 
$m_u:=v Y_u/\sqrt{2}$, $m_d:=v
Y_d/\sqrt{2}$, $m_e:=v Y_e/\sqrt{2}$ and $m_{\nu_D}:=v Y_\nu/\sqrt{2}$.
The final explicit term in (\ref{diracMass}) incorporates the see-saw
mechanism via a 6 by 6 complex
symmetric mass matrix. Since the elements in $m_{\nu_D}$ are much less than
those in $M$, one performs a rotation to obtain a 3 by 3 complex symmetric
mass matrix for the {\em light}\/ neutrinos. To a good approximation, these
coincide with the left-handed weak eigenstates ${\bm \nu'_L}$, whereas
three heavy neutrinos approximately correspond to the right-handed weak
eigenstates ${\bm \nu'_R}$. The neutrino mass term of (\ref{diracMass}) becomes, to a good
approximation, 
\begin{equation}
-  {\mathcal L}_{\nu} =
  \frac{1}{2} {\overline {\bm \nu_L'^c}} m_\nu {\bm \nu_L'} +
\frac{1}{2} {\overline {\bm \nu_R'^c}} M {\bm \nu_R'} + H.c., 
  \end{equation}
where $m_\nu:= m_{\nu_D}^T M^{-1} m_{\nu_D}$ is a complex symmetric 3 by 3
matrix. 

Choosing
$V_{I_L}^\dagger m_I  V_{I_R}$ to be diagonal, real and positive for $I
\in \{ u,d,e\}$, and
$V_{{\nu}_L}^T m_\nu  V_{{\nu}_L}$ to be diagonal,
real and positive 
(all in increasing order of mass
from the top left toward the bottom right of the matrix), we can identify the 
{\em non}-primed {\em mass}\/ eigenstates
\begin{eqnarray}
{\bf u_R}:= V_{u_R}^\dagger {\bf u_R}', \quad &
{\bf u_L}:= V_{u_L}^\dagger {\bf u_L}', \quad & 
{\bf d_R}:= V_{d_R}^\dagger {\bf d_R}', \nonumber \\
{\bf d_L}:= V_{d_L}^\dagger {\bf d_L}', \quad &
{\bf e_R}:= V_{e_R}^\dagger {\bf e_R}', \quad &
{\bf e_L}:= V_{e_L}^\dagger {\bf e_L}'. \nonumber \\
{\bm \nu_L}:= V_{\nu_L}^\dagger {\bm \nu_L}'.\quad & &
\label{fermion_rotations}
\end{eqnarray} 
We may then find the CKM matrix $V$ and the
Pontecorvo-Maki-Nakagawa-Sakata (PMNS) matrix $U$ in terms of the fermionic
mixing matrices:
\begin{equation}
V=V_{u_L}^\dagger V_{d_L}, \qquad U = V_{\nu_L}^\dagger V_{e_L}. \label{mix}
\end{equation}

\bibliographystyle{spphys}
\bibliography{b3l2}
\end{document}